\documentclass[conference]{IEEEtran}
\IEEEoverridecommandlockouts
\usepackage{cite}
\usepackage{amssymb,amsfonts}
\usepackage{amsmath}
\usepackage{algorithmic}
\usepackage[ruled, linesnumbered, noend]{algorithm2e}
\usepackage{graphicx}
\usepackage{textcomp}
\usepackage{xcolor}

\graphicspath{{figures/}} %Setting the graphicspath
\usepackage{comment}
\usepackage{caption}
\usepackage{subcaption}
\usepackage{float}
\usepackage{parskip}
\usepackage{afterpage} % for placing figures after a specific page

\newlength{\bibitemsep}
\newlength{\bibparskip}
\let\oldthebibliography\thebibliography
\renewcommand\thebibliography[1]{%
  \oldthebibliography{#1}%
  \setlength{\parskip}{\bibitemsep}%
  \setlength{\itemsep}{\bibparskip}%
}
\usepackage[left=0.625in,right=0.625in,top=0.75in,bottom=1in]{geometry}
\usepackage{soul}
\usepackage{soul}

\usepackage[font=scriptsize,labelfont=bf]{caption} % add this in preamble

\begin{document}

%\title{DRL-based Neighbor Discovery and Exposure Management for Swarming Directional UAVs}

\title{DQN-Driven Adaptive Neighbor Discovery for Directional Aerial Networks
\thanks{\textcopyright \enspace 2026 IEEE. Personal use of this material is permitted. Permission from IEEE must be obtained for all other uses, in any current or future media, including reprinting/republishing this material for advertising or promotional purposes, creating new collective works, for resale or redistribution to servers or lists, or reuse of any copyrighted component of this work in other works.}
\thanks{This work is supported in part by a grant from the U.S. Air Force Research Lab and NSF awards 2115215 and 2006683.}
\thanks{Distribution A. Approved for public release: Distribution Unlimited: AFRL-2025-5562 on 09 Dec 2025.}
}

% “Adaptive Transceiver Selection for Covert Connectivity in Directional Mobile Networks Using Deep Q-Learning”

% “Deep Reinforcement Learning-Based Transceiver Selection for Covert Directional Neighbor Discovery”

% “Balancing Connectivity and Covertness in Directional Mobile Networks via Deep Q-Networks”

% “DQN-Driven Adaptive Beam Probing for Covert Tactical and Aerial Networks”

% “Autonomous Directional Neighbor Discovery with Limited Exposure via Deep Q-Networks”

%\title{Eavesdropper-Avoiding Neighbor Discovery and Reachability for Multi-Sector Directional Wireless Systems via Deep Reinforcement Learning} 
% \thanks{This work is supported in part by NSF award 2115215 and US Air Force Research Lab.}
% \thanks{Distribution A. Approved for public release: Distribution unlimited: AFRL-2024-0317 on 19 Jan 2024.}
% }

\author{Md Asif Ishrak Sarder$^{1}$, and Murat Yuksel$^{1}$
and Elizabeth Bentley$^{2}$\\
{\small$^{1}$\textit{University of Central Florida, Orlando, FL, USA}}\\
{\small$^{2}$\textit{Air Force Research Lab, Rome, NY, USA}}\\
{\footnotesize MdAsif.IshrakSarder@ucf.edu, murat.yuksel@ucf.edu,
elizabeth.bentley.3@us.af.mil}
\vspace{-6mm}
}

\maketitle

% \begin{IEEEkeywords}
% Neighbor Discovery, Probability-of-Intercept, Directional Wireless, Super-6 GHz
% \end{IEEEkeywords}

\IEEEpeerreviewmaketitle

\begin{abstract}

Directional antenna systems are gaining substantial traction for aerial networks due to their higher gain, extended transmission range, and enhanced security. However, the requirement of beam alignment makes the task of finding and reaching neighbors challenging, particularly in a mobile setting. 
%Notable efforts have been made to find solutions towards optimizing beam configuration to facilitate connectivity. 
For wireless networks, privacy concerns play an equally critical role.
However, the problem of ensuring \textit{network-wide connectivity} while maintaining \textit{limited exposure} when probing around is still unexplored. We address this trade-off by proposing an adaptive transceiver selection protocol based on the Deep Q-Network (DQN) framework. Each node acts as an independent DQN agent and interacts with the environment to learn how to balance the trade-off. Since the directional nodes operate only based on local observations, we adopt a weighted mechanism that guides them in prioritizing either high reachability or privacy by adaptively tuning the probing patterns. 
Results show that DQN framework surpasses the Random and Q-Learning baselines. Weights favoring discovery provide higher probing efficiency and reachability, while weights prioritizing privacy ensure limited exposure at the cost of low reachability, eventually attaining higher objective value.
\end{abstract}

\section{Introduction}

Directional antenna systems are rapidly becoming central to next-generation wireless networks operating at super-6 GHz bands. The ability to provide very high throughput with wider bandwidth, spatial reuse, extended transmission range, and enhanced security with interference mitigation makes them particularly appealing for mobile and aerial platforms \cite{fanet, reza2024uav}. However, this very directionality with its narrow beam pattern brings unique challenges when it comes to ensuring network-wide connectivity for mobile self-configuring networks \cite{gossip}. Failure to precisely align the directional beam between nodes can cause a significant drop in signal quality 
%for mobile networks 
using very high frequencies like millimeter-wave (mmWave) or Terahertz (THz) bands \cite{mmwaveLoss}. 
In a highly dynamic environment, the tasks of neighbor discovery and reachability (nodes being able to reach each other) become time-variant and complex.
Further, privacy is an equally vital aspect that often gets overlooked. 
Since each node must probe continuously in different directions to stay connected with its peers, probe messages can easily be intercepted by unauthorized users and thus compromise Unmanned Aerial Vehicular (UAV) networks. 
Hence, a more sophisticated approach is required to ensure network-wide connectivity while maintaining privacy (by minimizing the exposure in the search space) from being overheard by undesired users.

\vspace{-2mm}
Prior efforts either treated the directional neighbor discovery and privacy separately \cite{pfl},\cite{regulateTxP_AN} or studied them jointly in static scenarios \cite{icccn}. 
%In our previous effort \cite{icccn}, we highlighted the research gap in addressing both fast neighbor discovery and covertness simultaneously in a directional setting. 
However, these are not sufficient to address the need for a mobile swarm of nodes that need to stay connected while maintaining privacy. In this paper, motivated by a UAV swarm responding to a disaster without any infrastructure support,
%a UAV swarm's need to of connectivity and covertness of UAVs in a
we seek to address the trade-off between \textit{high reachability} among the swarming nodes and \textit{limited exposure} to maintain privacy (or simply minimize the interference footprint of the swarm on others).

\vspace{-2mm}
We consider a fully decentralized mobile ad hoc network where directional nodes are configured with 2D multi-sector directional antenna systems, where each sector contains a directional transceiver. Each node probes through different sectors electronically \cite{elec_steer} to find neighbors within its line-of-sight (LoS). We adopt a Deep Reinforcement Learning (DRL) framework to realize an adaptive transceiver selection protocol, where each directional node, acting as an independent (autonomous) Deep Q-Network (DQN) agent, adaptively learns how to tune its probing patterns to strike a balance between the two conflicting goals by interacting with the environment. We introduce a weighted system that controls the probing patterns of the directional nodes based on individual observations, and thus guides them toward prioritizing either goals.
Key insights and contributions of this work include:
% \vspace{-5mm}
\begin{itemize}
\item Successfully conducting neighbor discovery to achieve network-wide connectivity using only one
directional transceiver rather than using multiple or all transceivers to send probe messages, without having any prior knowledge of positioning and channel state information (CSI) of the neighboring nodes.

\item Implementation of a model-free DQN framework for adaptive transceiver selection, where independent DQN agents learn to balance the trade-off between high reachability and limited exposure for privacy concerns by adaptively changing probing directions under mobility.

\item A weighted mechanism that enables the directional nodes to dynamically modify their probing patterns to prioritize either goals between exploring different directions and limited exposure based on local observations.
\end{itemize}

\section{Related Works}

Shifting away from the traditional omni-directional antenna systems, directional wireless systems have demonstrated notable performance improvements.
By leveraging transceivers with the ability to mechanically \cite{mecha_steer} or electronically \cite{elec_steer} steer the beam direction, existing neighbor discovery algorithms can schedule the beam configuration either in a randomized \cite{gossip}, deterministic \cite{nd_rw1}, or probabilistic manner \cite{icccn}.
To expedite discovery completion with reduced interference, several strategies have been adopted: supervised scheduling where receivers switch the listening direction while the transmitter keeps probing at the same direction \cite{nd_rw1}, gossip-based hybrid discovery where nodes exchange information about their partially known topology \cite{gossip}, and joint utilization of main and side lobes to map the neighboring nodes from received signals \cite{side_lobe}. 
However, factors like packet collisions and link security are often overlooked.  

In an effort to design an intelligent beam configuration-based protocol via reinforcement learning (RL), \cite{basic_q_nd} uses Q-learning (QL) to develop a time-synchronized protocol to achieve a higher discovery rate. 
In a distributed multi-agent setup, \cite{q_collision} considers collision feedback as a reward signal for QL. This approach exploits the intuition that a collision observed at a certain direction indicates neighbors being in that direction, making it worthwhile to revisit that same sector.
% However, under half-duplex operation, these methods did not fully utilize both the transmission and reception mode observations. 
\cite{erttnd} proposes an RL-based two-way transmit–receive discovery algorithm that allows learning from information gathered in both transmission and reception modes.
Beyond tabular RL algorithms, \cite{pfl} uses federated learning 
%integrates Personalized Federated Learning (PFL) 
with Deep Deterministic Policy Gradient (DDPG), reporting faster completion while addressing the scarcity and heterogeneity of per-node data. 
However, robust decentralized aggregation of local models remains a challenge due to unstable connectivity among mobile nodes.

Using directional antennas, common physical layer security  strategies against being overheard by undesired neighbors involve careful regulation of transmit power \cite{regulateTxP}, to reduce the probability-of-intercept (POI). Optimizing the transmit power while adding artificial noise \cite{regulateTxP_AN} or adapting the UAV trajectory \cite{fanet} shows additional privacy capabilities. However, varying transmit power on-the-fly adds complexity \cite{nd_rw1}. 

Our approach is different from these prior efforts in terms of avoiding the need to have physical layer information (which is impractical in wireless ad hoc deployments) and takes a minimalist approach, i.e., assumes no GPS and no coordination among the swarming nodes.

% \cite{regulateTxP} proposes optimizing the transmit power to achieve reduced beam exposure and low probability-of-intercept (POI). 
% In \cite{regulateTxP_AN}, coupling Artificial Noise with optimized transmit power has shown effectiveness tackling hybrid eavesdroppers under imperfect CSI condition. 
% Further, authors in \cite{fanet} propose joint optimization of trajectory and TP via Multi-Agent DDPG algorithm, coupled with friendly-jammer UAVs injecting optimized AN, for a Flying Ad Hoc Network that attains high throughput under covertness constraints.
% However, varying TP on-the-fly can add more complexity to the scheme \cite{nd_rw1}. 
% Also, some methods often make assumption of having detailed information about eavesdroppers, which is somewhat impractical in wireless ad-hoc deployments. 

% \vspace{-1mm}
\section{System Model}

We consider a set of directional wireless nodes $\mathcal{N}$ and \textit{M} omni-directional undesired users deployed in 2-D geographical plane. Considering a fully decentralized environment with no prior coordination, the directional nodes are expected to self-configure the network while maintaining privacy.
% \subsection{Assumptions}
All directional nodes are homogeneously configured with an electronically steerable multi-sector antenna system that allows full-duplex operation \cite{icccn}. Each sector is equipped with a directional transceiver, collectively providing $360^{\circ}$ coverage in the node's search space. Each node is assigned a unique identifier (e.g., a MAC address). All the directional nodes and undesired users are mobile in the air. Directional nodes start probing without having any idea about their surroundings at initial stages. Further, they are oblivious to the presence of undesired users and to the total node count in their swarm.

\subsection{Channel Model}
To establish an LoS link between nodes $N_{i},N_{j} \in \mathcal{N}$, both $N_{i}$ and $N_{j}$ must lie within each other's transmission ranges ($R$) and field-of-views ($FoV$), and the received powers at each end i.e., $\mathcal{P}_r(i,j)$ and $\mathcal{P}_r(j,i)$ must exceed a minimum threshold, $\mathcal{P}_o$. The received power $\mathcal{P}_r(i,j)$ at receiver $N_j$ from transmitter $N_i$ is calculated by \cite{Pr_eqn}:
\begin{equation}\label{pr_eqn}
\mathcal{P}_r(i,j) = \mathcal{P}_t k_0 G_i^t(\theta_i - \phi_{i\xrightarrow{}j}) G_j^r(\theta_j - \pi - \phi_{i\xrightarrow{}j}) l_{ij}^{-\eta},
\end{equation}
where $\mathcal{P}_t$ is the transmission power of node $N_i$, 
$k_0=(\frac{\lambda}{4\pi})^2$ is the free-space path loss factor with $\lambda$ being the carrier signal wavelength, $G_i^t(\cdot)$ and $G_j^r(\cdot)$ are the directional antenna gain functions of the transmitter $N_i$ and receiver $N_j$, respectively. 
Further, $\theta_i$ and $\theta_j$ are the beam divergence half-angles of $N_i$ and $N_j$, respectively. 
$\phi_{i \to j}$ denotes the angular deviation between nodes $N_i$ and $N_j$, which is determined by the orientations and relative position of $N_i$ and $N_j$.
$l_{ij}$ denotes the Euclidean distance between $N_i$ and $N_j$, and $\eta$ is the path loss exponent.
We assume a sector's angular width is equal to $FoV$ and the beamwidth covers the entire sector, i.e., $\theta=\frac{FoV}{2}$.

\subsection{Directional Antenna Model}

We adopt the IEEE 802.15.3c standardized directional antenna model \cite{antenna_gain} to characterize the directional gain patterns, prioritizing only the main lobe of the antenna beam pattern, since we focus on LoS links only.
% Only the main lobe of antenna pattern is prioritized as the side lobes can be safely overlooked for wireless systems operating at very high frequencies \cite{side_lobe_off}. 
The gain of such a directional antenna with divergence angle $\theta$ is stated in terms of decibels (dB) as:
\begin{equation}\label{eqn2}
    G(\theta) = G_0 - 3.01 \cdot \left(\frac{2\theta}{\theta_{\text{-3dB}}}\right)^2,\quad\ 0^{\circ} \leq \theta \leq \frac{{FoV}}{2}
\end{equation}
where 
%$\theta$ is an arbitrary angle within the range [$0^{\circ},\frac{{FoV}}{2}$], 
$\theta_{\text{-3dB}} = FoV/2.6$ is the half-power beamwidth (HPBW) angle in degrees and 
$G_0 = 10\log{\left(1.6162 / \sin\left(\theta_{\text{-3dB}}/2\right)\right)^2}$
%$G_0 = 10\log{\left(\frac{1.6162}{\sin\left(\frac{\theta_{\text{-3dB}}}{2}\right)}\right)^2}$
is the maximum antenna gain in dB.

\subsection{Swarm Mobility Model}
%Inspired by the swarm behavior observed in aerial networks , 
We adopt a swarm-coordinated mobility model \cite{mobility_model} for $\mathcal{N}$ directional nodes while $M$ undesired users move randomly within the network area. The initial deployment of $\mathcal{N}$ nodes follows a grid layout to ensure uniform distribution.
Inside the swarm, a coherent global trajectory is maintained by the directional nodes. Further, they simultaneously perform random movement, with velocity $v$, within a circular roaming zone of radius $R_{roam}$ centered around individual swarm-aligned positions. 
The global trajectory is controlled by a random drift vector that gets inverted as soon as a node approaches the network boundary. 
The local mobility of the directional nodes is constrained by: \textit{Local Boundary Check} where the node reflects off the wall when it hits the boundary of the roaming zone, and \textit{Global Boundary Check} where the node reflects off the network boundary to ensure it stays within the valid network area.  
In this way, even though the density is uniform inside the swarm, random local movement can still induce transient variations in neighborhood proximity that temporarily create denser or sparser regions.

\subsection{Neighbor Discovery and Privacy}
To self-configure the network, we assume each node periodically enters \textit{Neighbor Discovery Mode}. During each NDM interval, a node uses only one of its on-board $K$ sectors to send a directional probe message which could resemble Synchronization Signal Blocks (SSBs) of 3GPP \cite{3GPP-NR}. Once a handshake is completed between two nodes, a communication link is established. A node is limited to finding at most 1 neighbor per NDM interval, while receiving more than 1 probe message is considered a collision. We assume nodes have the capability of perceiving a collision event. 
%\noindent \textit{Eavesdropping:} 
% During each NDM interval, an eavesdropper is limited to detect at most 1 communication link. In that case, both participating directional nodes must fall within the detection range of the eavesdropper, denoted by $R_d$.
During each NDM interval, an undesired user is limited to detecting at most 1 directional node that has probed toward the undesired user while falling within the undesired user's detection range, $R_d$.

\subsection{Link Maintenance and Reachability}
In a mobile network, neighbor discovery alone can become insufficient to ensure the validity of all active links. For nodes $N_i, N_j, N_k \in \mathcal{N}$, let's assume $N_i$ finds $N_j$ at NDM interval $t$. If $N_i$ finds a different node $N_k$ at the next interval, this does not guarantee that the prior link between $N_i$ and $N_j$ remains active -- which calls for a separate link management framework to assess the temporal validity of the active links. In our setting, a lightweight timeout-based link maintenance method is adopted that operates solely based on probing outcomes. 
Once node $N_i$ links with $N_j$, a \textit{Link Alive Counter} is initiated. The counter gets incremented whenever the same neighbor fails to be re-detected during the next NDM interval, otherwise resets to zero upon reaffirmation. If the counter exceeds a predefined timeout threshold, the link expires.

Once link validations are complete, let $\mathcal{L}(t)$ record all active links at instance $t$. For $N=|\mathcal{N}|$, in an undirected graph $G(t) = (N, \mathcal{L}(t))$,  two nodes $N_i,N_j \in \mathcal{N}$ can be connected either directly via active link $L_{ij} \in \mathcal{L}(t)$ or indirectly through immediate relay nodes using multi-hop routing.
The instantaneous network topology at NDM interval $t$ can be represented by an adjacency matrix $A(t)_{N \times N}$, where:
\begin{equation}
A_{ij}(t) = 
\begin{cases}
1, & \text{if $L_{ij} \in \mathcal{L}(t)$},\\
0, & \text{else}.
\end{cases}
\end{equation}
From $A(t)$, we construct an undirected graph $G(t)$. 
For node $N_i$, let $\mathcal{C}_i(t)$ be the set of unique nodes reachable from $N_i$ via direct or multi-hop links in $G(t)$. Then, the reachability of node $N_i$ is defined as:
\begin{equation}
r_i(t) = \frac{|\mathcal{C}_i(t)|}{N-1}, \quad 0\leq r_i(t) \leq 1
\end{equation}
where $\left| \cdot \right|$ is the cardinality of the set and $N-1$ denotes total possible neighbors of $N_i$ across the network.

% The network-wide reachability $R_{net}(t)$ at NDM interval $t$ can be computed by:
% \begin{equation}
% R_{net}(t) = \frac{1}{N} \sum_{i=1}^N r_i(t).
% \end{equation}

\section{Problem Formulation}

To discover and maintain communication in a decentralized setup, nodes must adapt their beam configuration based on local observations -- making this a transceiver/sector selection problem for the following NDM interval(s).
Since the total node count is unknown to directional nodes, computing reachability locally is no longer viable. 
Discovery likelihood acts as a good estimator of reachability.
% as high discovery rate can ensure a stable reachability with peers.
Let $\mathcal{K} = \{1, 2,..., K\}$ be the set of sectors a node will select from, $p_k^n(t)$ indicates the probability of finding a neighbor at sector $k$, and $p_k^e(t)$ denotes the probability of an undesired user existence at sector $k$ at NDM interval $t$. Let $\varepsilon(t) \in \mathcal{K}$ be the sector used for probing at NDM interval $t$. 
% Let $r_i(t) \in [0,1]$ represent the reachability of a node $N_i$ given the instantaneous network topology. 
% We assume that the past probing outcomes are available for a window of $W$ intervals. 
% Let $\Delta r_{i,k}(t)$ be the net reachability gain using sector $k$ at NDM interval $t$ at node $N_i$, i.e., $\Delta r_{i,k}(t) = r_i(t) - r_i(t-1)$. 
Then the problem of selecting the next sector to be probed at node $N_i$ for the \textit{next interval only} can be written as:

\vspace{-2.0em}
% % {\small
% \begin{IEEEeqnarray}{lCl}
% \textbf{(P):} 
% & 
% % \hspace{-0.1em}
% \max_{\varepsilon(t+1)} &
% p_{i,\varepsilon(t+1)}^{n}(t+1)
% - \!\!\sum_{j=t-W+1}^{t+1}\! p_{i,\varepsilon(j)}^{e}(j) \nonumber
% , \label{sector-selection-problem}\\[3pt]
% && \text{such that} \nonumber\\[1pt]
% && \varepsilon(j)\!\in\!\mathcal{K},\; j=[t-W+1,..,t+1], 
% \label{eq:constraint1}\\
% && |\varepsilon(j)|\!\le\!1,
% \label{eq:constraint2}\\
% && 0\!\le\!p_{i,\varepsilon(t+1)}^{n}(t\!+\!1)\!\le\!1, \text{and}
% \label{eq:constraint3}\\
% && 0\!\le\!p_{i,\varepsilon(j)}^{e}(j)\!\le\!1,\;
%    j=[t-W+1,..,t+1]. 
% \label{eq:constraint4}
% % && -1\!\le\!\Delta r_{i,\varepsilon(j)}(j)\!\le\!1,\;
% %    j=[t-W+1,..,t+1] .
% % \label{eq:constraint5}
% \end{IEEEeqnarray}
% % }
% {\small
\begin{IEEEeqnarray}{lCl}
\textbf{(P):} 
& 
% \hspace{-0.1em}
\max_{\varepsilon(t+1)} &
p_{i,\varepsilon(t+1)}^{n}(t+1) - p_{i,\varepsilon(t+1)}^{e}(t+1) \nonumber
, \label{sector-selection-problem}\\[3pt]
&& \text{such that} \nonumber\\[1pt]
&& \varepsilon(t+1)\!\in\!\mathcal{K}, \quad\text{and}
\label{eq:constraint1}\\
&& |\varepsilon(t+1)|\!\le\!1.
\label{eq:constraint2}
% && 0\!\le\!p_{i,\varepsilon(t+1)}^{n}(t\!+\!1)\!\le\!1, \text{and} 
% \label{eq:constraint3}\\
% && 0\!\le\!p_{i,\varepsilon(t+1)}^{e}(t\!+\!1)\!\le\!1.
% \label{eq:constraint4}
% && -1\!\le\!\Delta r_{i,\varepsilon(j)}(j)\!\le\!1,\;
%    j=[t-W+1,..,t+1] .
% \label{eq:constraint5}
\end{IEEEeqnarray}
% }
Here, 
% the first term aggregates the net reachability gains achieved over the window $W$, 
% the middle term 
the first term is the probability of finding a neighbor in the following interval, while the last term is the probability of the probe messages being overheard by undesired users that quantifies the POI. The first term captures the expected gain: the higher discovery likelihood, the steadier or even upward trend for the reachability gain for the following interval. Baseline objective is the maximization of the difference.
Constraints (\ref{eq:constraint1}) and (\ref{eq:constraint2}) make sure either none or at most one sector is selected for probing for any NDM interval. 
% Constraints (\ref{eq:constraint3}) and (\ref{eq:constraint4}) satisfy that the probabilities are in [0,1]. 
% Since mobility can break links and drop connectivity, $\Delta r$ can be negative; hence, $\Delta r$ is bounded to [-1,1] by constraint (\ref{eq:constraint5}).

% [from ICCCN] --Two aspects make the problem (P1) difficult. \textit{First}, the probabilities $p_v^n(.)$ and $p_v^e(.)$ are unknown and time-variant due to external factors such as mobility. \textit{Second}, the probability of finding a new neighbor in future depends on the past sector selections, i.e., $p_v^n(t+1)$ depends on the sectors probed and the neighbors found in the past. In other words, $p_v^n(t+1) \thicksim \epsilon(j), j=[t-W+1,..,t]$ and $p_v^n(t+1) \thicksim \mathcal{N}_k(j), j=[t-W+1,..,t]$. Practical solutions lie in skillfully estimating these unknowns in real time, for which we provide a method in this effort. 

% One way to connect the two is to explain that the probabilities in this problem formulation are not readily available. Hence, we have to come up with a way to approximate them from local observations.

Under mobility, the probabilities $p_{t+1}^n(.)$ and $p_{t+1}^e(.)$ become unknown and time variant.
Instead, directional nodes can use recent probing records to estimate these unknowns in real time, as past sector selections heavily influence the discovery likelihood in the near future. 
Excessive exploration may improve discovery likelihood and reachability, but also hurt privacy by causing broader exposure in the search space. 
% This motivates to introduce a weighted mechanism that acts as a knob between high reachability and staying covert by adaptively tuning the probing patterns. 
This motivates to introducing a \textit{weighted mechanism} that 
% addresses this trade-off and  
allows a directional node to self-tune its probing patterns according to network dynamics. 
% Gaining higher probing efficiency while linking up with unique neighbors indicates node's drift in denser region: frequent probing direction change can help increasing the reachability. 
% Inversely, a drop in probing efficiency points to its drift towards sparser region. In that case, reusing a subset of sectors with recent success can preserve connectivity with reduced exposure. 
Gaining higher probing efficiency in denser region can allow for frequent probing direction change to improve reachability.
Conversely, lower efficiency in sparser region can push towards reusing a subset of sectors with recent success can preserve connectivity with reduced exposure. 
Practical solutions lie in skillfully tuning the weights associated with each goal.

\begin{figure*}[htb]
    \centering        
    \includegraphics[width=1\linewidth]{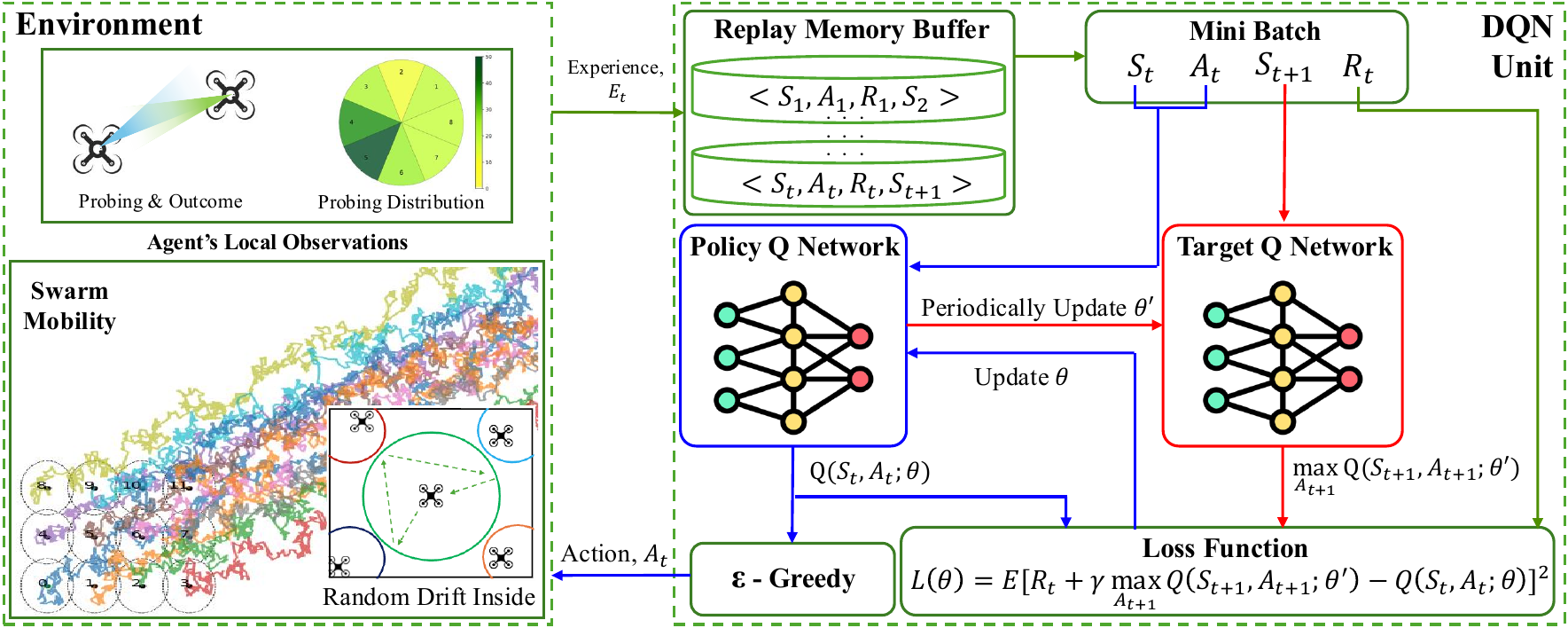}
    \caption{The block diagram of Deep Q-Network (DQN) framework}
    \label{fig:dqn_setup}
    \vspace{-6mm}
\end{figure*}

\section{DRL for Adaptive Transceiver Selection}
% \subsection{Motivation}
% For the problem in (\ref{sector-selection-problem}) 
For problem (P) that couples two conflicting goals, it is difficult to find a closed-form solution since the environment is non-stationary and each node only has a partial observability of the entire network gathered from recent probings. 
%Further, a decentralized network with nodes lacking a birds-eye view of the entire instantaneous topology at decision time makes a centralized optimization approach impractical. 
Therefore, we adopt a model-free DRL framework i.e., Independent Deep Q-Network (IDQN) algorithm \cite{IDQN}, 
where each directional node operates as an autonomous DQN agent, independently making its own decisions to optimize its transceiver selection policy based on individual observations. The main idea is to turn the global objective of maximizing reachability with limited exposure into a per-node decision making process driven by locally accessible information. Fig.\ref{fig:dqn_setup} provides a block diagram of our proposed DQN framework. 

% Therefore, we adopt a fully distributed, model-free approach where each nodes learns its own transceiver selection policy from local observation made over recent intervals only. The main idea behind is to turn the global objective of maximizing reachability with limited POI into a per-node decision making process driven by locally accessible information.
% % This paper employs Deep Q-Learning (DQN) framework since it is well-suited to handle discrete action space, where outcomes are sparse and explicit modeling of environment dynamics is not required.

% % \subsection{Overview of Deep Q-Learning}

% \subsection{Proposed Independent Deep Q-Learning Framework}
% This paper proposes a distributed Deep Reinforcement Learning framework i.e., Independent Deep Q-Learning (IDQN) where each directional node operates as DQN agent independently making its own decisions based on individual observations in a multi-agent setting \cite{IDQN}. Using the past probing records as local observations, nodes can quantify the probing efficiency and corresponding exposure in its search space.

\noindent \textbf{States:}
For a DQN agent with $K$ on-board transceivers and a memory system that can store recent probings of length $W$, the local state vector $S_t$ at NDM interval $t$ is the concatenation of three different components:

\noindent \textit{1. Probing History:} Let $S_A$ be the vector logging all actions taken by the agent over the last $W$ intervals for probing:
\vspace{-1mm}
\begin{equation}
S_A = [A_{t-W+1}, A_{t-W+2},..., A_{t}] \label{eq:probing_history}.
\end{equation}
where $A_\tau \in \{1,2,..,K\}$ denotes the sector probed at NDM interval $\tau$. It maintains a First-In-First-Out (FIFO) structure to log the latest data while removing the oldest one.
% \textcolor{red}{[ref]}.

\noindent \textit{2. Outcome History:}
Let $S_o$ be the FIFO vector logging all the outcomes of the last $W$ probings:
\begin{equation}
S_o = [o_{t-W+1}, o_{t-W+2},.., o_{t}] = \{0,0.5,1\}^W \label{eq:probing_outcome}.
\end{equation}
where $o_\tau \in \{0, 0.5, 1\}$ at NDM interval $\tau$ denotes three possibilities: either no discovery at all, or a collision is observed, or a neighbor is discovered, respectively. 
%It also uses FIFO structure to log latest probing outcomes.

\noindent \textit{3. Probing Distribution:}
To observe the frequency of probing different sectors over the last $W$ intervals, we utilize the probing history $S_A$ outlined in (\ref{eq:probing_history}).
% \noindent For each sector $k \in K$, the count of total uses over a window can be defined by:
% \vspace{-1mm}
\noindent Let $C_W = \left[C_1, C_2,..,C_K\right]$ be the vector logging the total probe count for each sector $k \in K$ over the last $W$ intervals, where:
\begin{equation}
C_k = \sum_{\tau = t-W+1}^t f\left(A_{\tau},k\right),  
\quad f\left(A_{\tau},k\right)= 
\begin{cases}
1, & A_\tau = k,\\
0, & A_\tau \neq k.
\end{cases}
\end{equation}
$C_W$ is then normalized, leading to the probing distribution of different sectors, denoted by $P_t = \left[P_{t,1},P_{t,2},..,P_{t,K}\right]$, that is computed at NDM interval $t$ as:
\vspace{-5mm}
\begin{align}
P_{t} = \frac{C_W}{W},  \quad \sum_{k=1}^K P_{t,k} = 1 \label{eq:dist}.
\end{align}
\vspace{-2mm}
% \noindent Total count for each sector $k$ is normalized in (\ref{eq:norm}), leading to the distribution of per-sector use $p_t$ computed for NDM interval $t$ in (\ref{eq:dist}):
% \begin{align}
% p_{t,k} &= \frac{c_{t,k}}{W}, \quad \text{for $k = 1,.., K$} \label{eq:norm}\\ 
% p_t &= (p_{t,1},..,p_{t,K}) \in [0,1]^K, \quad \sum_{k=1}^K p_{t,k} = 1 \label{eq:dist}
% \end{align}

% Then the state vector and its dimension are:
% \begin{equation}
% s_t = [s_t^a \parallel s_t^o \parallel p_t] \in \mathbb{R}^d
% \end{equation}
\noindent \textbf{Actions:}
At each NDM interval $t$, a DQN agent uses $\epsilon$-greedy algorithm to choose one sector to probe for the following interval. So, the action space $A_t$ at interval $t$ can be:
\begin{equation}
A_t \in \mathcal{A} \triangleq \{1,2,..,K\}.
\end{equation}
\textbf{Weighted Objective Function:}
The probing efficiency (PE) $O_p$ over a window $W$ can be computed using recent probing outcomes $S_o$:
\begin{equation}
% O_t^p = \frac{1}{W} \sum_{\tau=t-W+1}^t o_\tau \in [0,1]
O_p(t) = \frac{1}{W} \cdot \sum_{\tau=t-W+1}^t o_\tau, \text{ where } o_\tau \in S_o.
\end{equation}
% where $o_\tau \in \{0,1\}$ indicates probing success at interval $\tau$. 
% High $O_p$ ensures a stable if not higher reachability with its peers.
% Higher probing efficiency ensures a stable if not higher reachability with its peers.

% From the probing distribution $P_t$ outlined in (\ref{eq:dist}), corresponding mean $\mu$ and standard deviation $\sigma$ at interval $t$ can be calculated by:

The standard deviation (STD) $\sigma$ of the probing distribution $P_t$ at NDM interval $t$ can be calculated by:
\begin{align}
% \mu &= \frac{W}{k} \\
\sigma_p(t) &= \sqrt{\frac{1}{K} \cdot \sum_{k=1}^K \left(P_{t,k} - \mu \right)^2}.
\end{align}
where $\mu = \frac{W}{K}$ 
% $\mu = \frac{\sum_{k = 1}^K C_k}{K}$
% $\mu = \sum_{k = 1}^K C_k/{K}$
is the mean of $P_t$.

\noindent For a uniform distribution where all the sectors are equally used, the corresponding STD, $\sigma_{min} = 0$. As for a skewed distribution where only one of the sectors is used constantly, we get:
\begin{equation*}
\sigma_{max} = \sqrt{\frac{1}{K} \cdot\left[\left(W - \mu\right)^2 + \left(K-1\right)\cdot\mu^2\right]}.
\end{equation*}
Using the values above, privacy $O_c$ is quantified by finding the normalized Coefficient of Variation (CV) of the recent probing distribution:
\begin{align}
O_c(t) = \text{CV}_{norm} = \frac{\sigma_p(t) - \sigma_{min}}{\sigma_{max} - \sigma_{min}}, \quad O_c(t) \in [0,1].
% O_c(t) = CV_{norm} = \frac{\mu \cdot[\sigma_p(t) - \sigma_{min}]}{\mu \cdot[\sigma_{max} - \sigma_{min}]}, \quad CV_{norm} \in [0,1].
\end{align}

We combine the two conflicting goals with a convex weight $w \in [0,1]$:
\begin{equation}
% \mathcal{O}_t = w \cdot O_t^p + (1-w) \cdot O_t^c.
O(t) = w \cdot O_p(t) + (1-w) \cdot O_c(t).
\label{eq:weighted_obj}
\end{equation}

\noindent\textbf{Reward:}
The reward quantifies the change in the exponentially weighted moving average (EWMA) of the objective $O(t)$, i.e., $\tilde{O}(t)$, because of the latest action $A_t$ at NDM interval $t$:
\begin{align}
\tilde{O}(t) &= \alpha\cdot O(t) + (1-\alpha) \cdot \tilde{O}(t-1), \\
\Delta(t) &= \tilde{O}(t) - \tilde{O}(t-1),\\
% \mathcal{R}_t &= 
R_t &= 
\begin{cases}
+1, & \Delta(t) > 0,\\
0, & \Delta(t) = 0,\\
-1, & \Delta(t) < 0.
\end{cases}
\end{align}
where $\alpha\in(0,1]$ is the smoothing factor. This aligns the RL signal with the goal: actions that increase the objective are rewarded; actions that hurt the balance are penalized. Hyper-parameters for DQN training are charted in Table \ref{tab:dqn-parameter}.
\begin{table}
\caption{DQN Hyper-parameters}
\label{tab:dqn-parameter}
\centering
\begin{tabular}{|c|c|}
% \begin{tabular}{@{}lccc@{}}
\hline 
\textbf{Parameter} & \textbf{Value}\\ \hline
Mini Batch Size & 128 \\ \hline
Replay Memory Buffer Size & 20000 \\ \hline
Hidden Layers & 128\_128\_128\_128 \\ \hline
Discount Factor$^*$ ($\gamma$) & 0.90 \\ \hline
Learning Rate$^*$ ($\alpha$) & 0.0003 \\ \hline
Max. Exploration Rate$^*$ & 1.0 \\ \hline
Min. Exploration Rate$^*$ & 0.35 \\ \hline
% Activation Function & ReLU \\ \hline
\end{tabular}
\\
$^*${\scriptsize Applies to the Q-Learning baseline as well.}
\end{table}

% \begin{table}[htbp]
% \caption{DQN Hyper-parameters}
% \label{tab:dqn-parameter}
% \centering
% \setlength{\tabcolsep}{4pt}% tighter padding
% \begin{adjustbox}{width=\columnwidth} % guarantees no overfull hbox
% \begin{tabular}{@{}>{\raggedright\arraybackslash}p{0.48\linewidth}ccc@{}}
% \toprule
% \textbf{Parameter} & \textbf{\textit{w}=0.1} & \textbf{\textit{w}=0.5} & \textbf{\textit{w}=0.9} \\
% \midrule
% Mini Batch Size & 128 & 128 & 128 \\
% Replay Memory Buffer Size & 5000 & 5000 & 5000 \\
% Discount Factor ($\gamma$) & 0.9 & 0.9 & 0.9 \\
% Learning Rate ($\alpha$) & 0.001 & 0.001 & 0.001 \\
% Maximum $\epsilon$ Rate & 1 & 1 & 1 \\
% Minimum $\epsilon$ Rate & 0.15 & 0.15 & 0.15 \\
% Hidden Layers & 128\_128\texttt{-}128 & 128\texttt{-}128\texttt{-}128 & 128\texttt{-}128\texttt{-}128 \\
% Activation Function & ReLU & ReLU & ReLU \\
% \bottomrule
% \end{tabular}
% \end{adjustbox}
% \end{table}

% \subsection{Pause and Resume Training}

\vspace{2mm}
\begin{table}
\caption{Simulation Parameters}
\label{tab:simulation-parameter}
\begin{center}
    \centering
    \begin{tabular}{|c|c|c|} \hline 
         \textbf{Parameter}&  \textbf{Symbol}& \textbf{Value}\\ \hline 
         \# Directional Nodes&  $N$& 12\\ \hline 
         \# Undesired Users &  $M$& 3\\ \hline 
         % Transmission Range for $N$ nodes&  $R$& 30 m\\ \hline 
         % Detection Range for $M$ nodes&  $R_d$& 10 m\\ \hline 
         \# On-board Transceivers&  $K$& 8\\ \hline 
         Field-of-View&   $FoV$& $360^{\circ}/K$\\ \hline 
         Transmission Power&  $\mathcal{P}_t$& 10 mW\\ \hline 
         Path Loss Exponent&  $\eta$& 2\\ \hline 
         Wavelength&  $\lambda$& $850\times 10^{-9}$ m \\ \hline
         Minimum Received Power&  $\mathcal{P}_o$& 0.5 mW\\ \hline
         Node Velocity&  $v$& 1 m/s\\ \hline
         Weights&  $w$& \{0.1, 0.5, 0.9\}\\ \hline
         - &$W$& 10 NDM Intervals\\ \hline
         Link Alive Threshold & - & 5 NDM Intervals \\ \hline
    \end{tabular}
\end{center}
\vspace{-4mm}
\end{table}

\begin{figure}
    \vspace{-3mm}
    \centering
    \begin{subfigure}[t]{0.49\linewidth}
        \centering
        \includegraphics[width=1\linewidth]{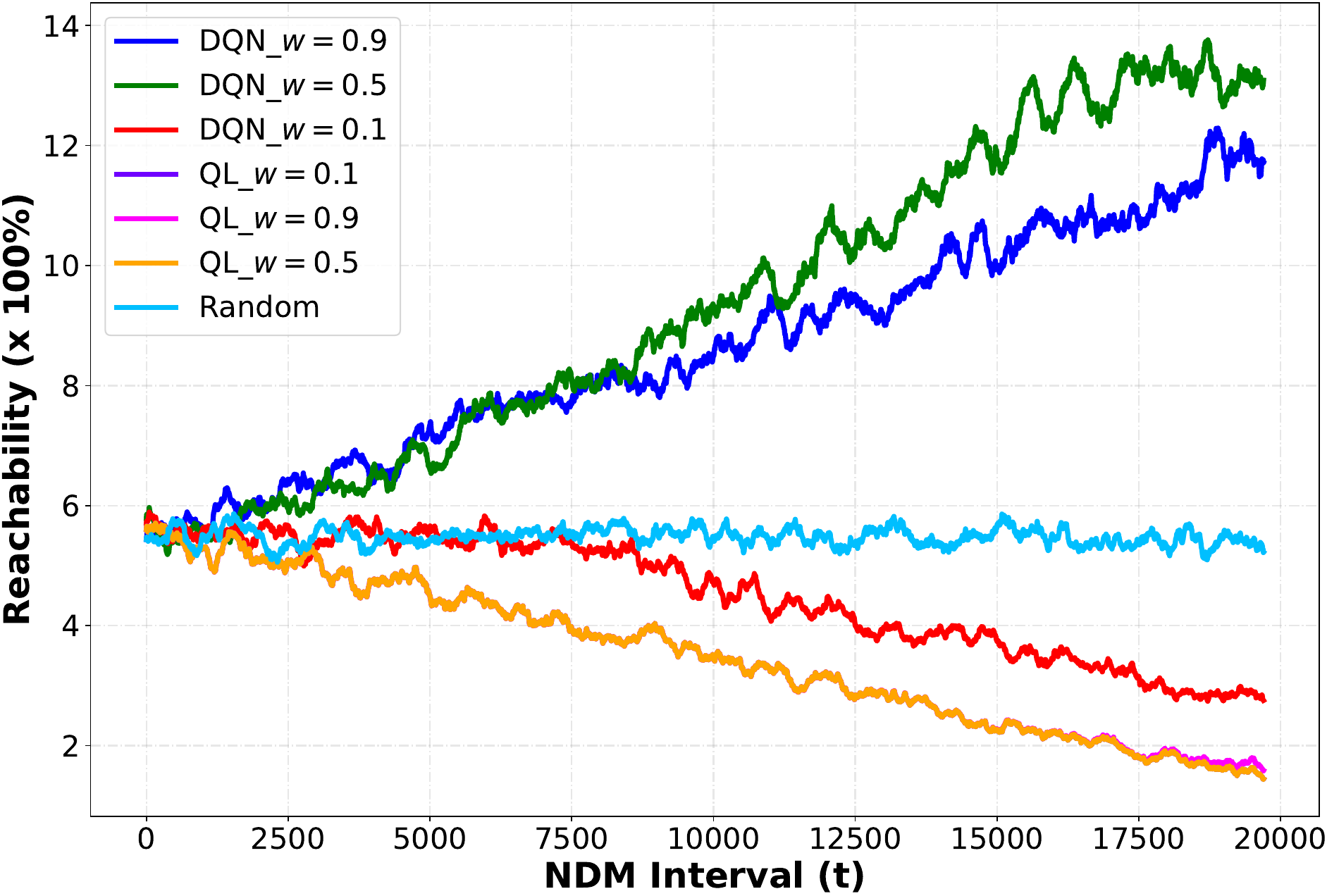}
        \captionsetup{format=hang}
        \caption{Reachability across the swarm over time}
        \label{alg_reach_comp}
    \end{subfigure}
    \begin{subfigure}[t]{0.49\linewidth}
        \centering
        \includegraphics[width=1\linewidth]{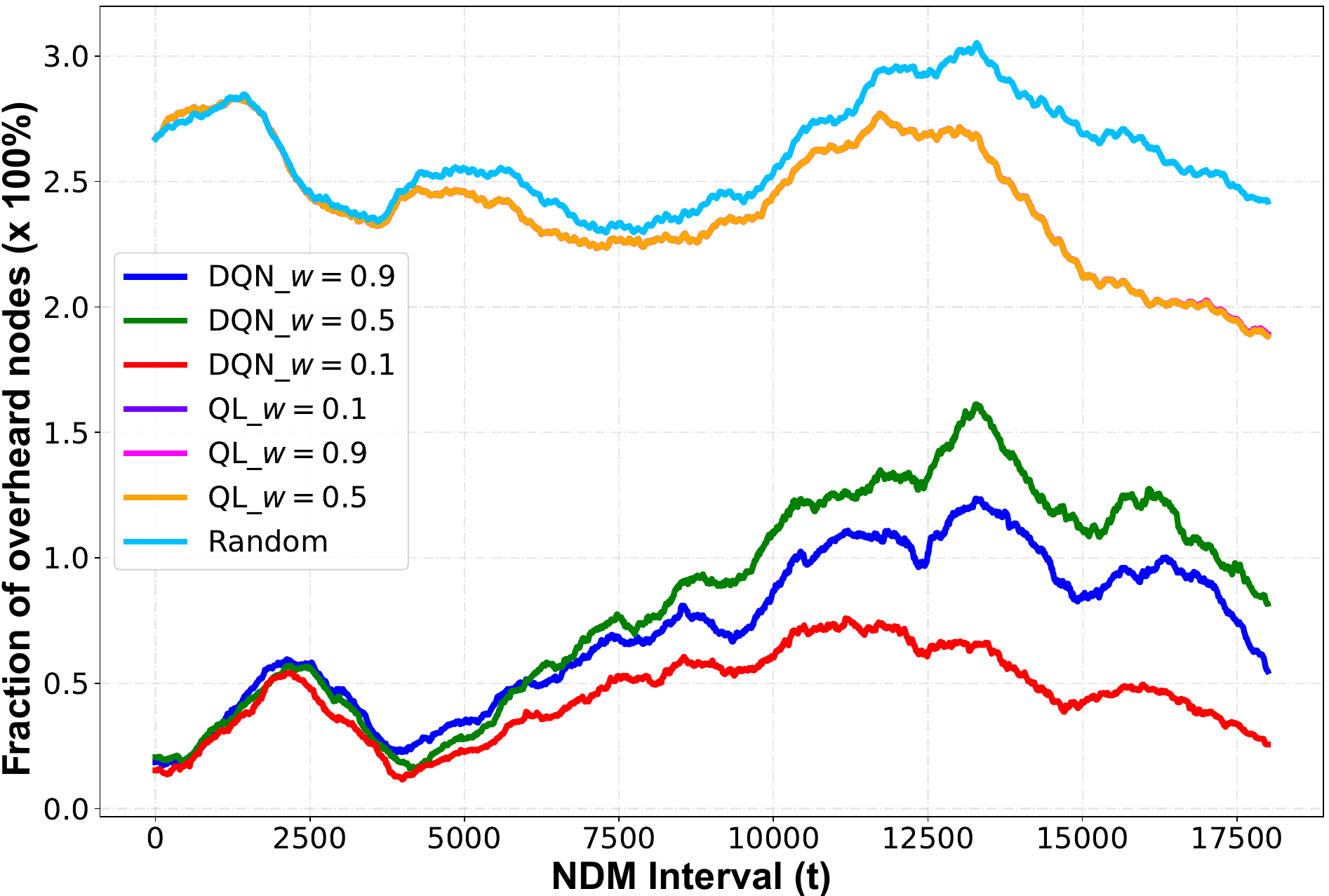}
        \captionsetup{format=hang}
        \caption{Fraction of nodes overheard by undesired users over time}
        \label{alg_eve_comp}
    \end{subfigure}
    \captionsetup{format=hang}
    \caption{Performance comparison of reachability and privacy breach occurrence between Random, Q-Learning, and DQN algorithms for different weights}
    \label{alg_reach_eve_comp}
\end{figure}

\begin{figure}
    \centering
    \begin{subfigure}[t]{0.49\linewidth}
        \centering
        \includegraphics[width=1\linewidth]{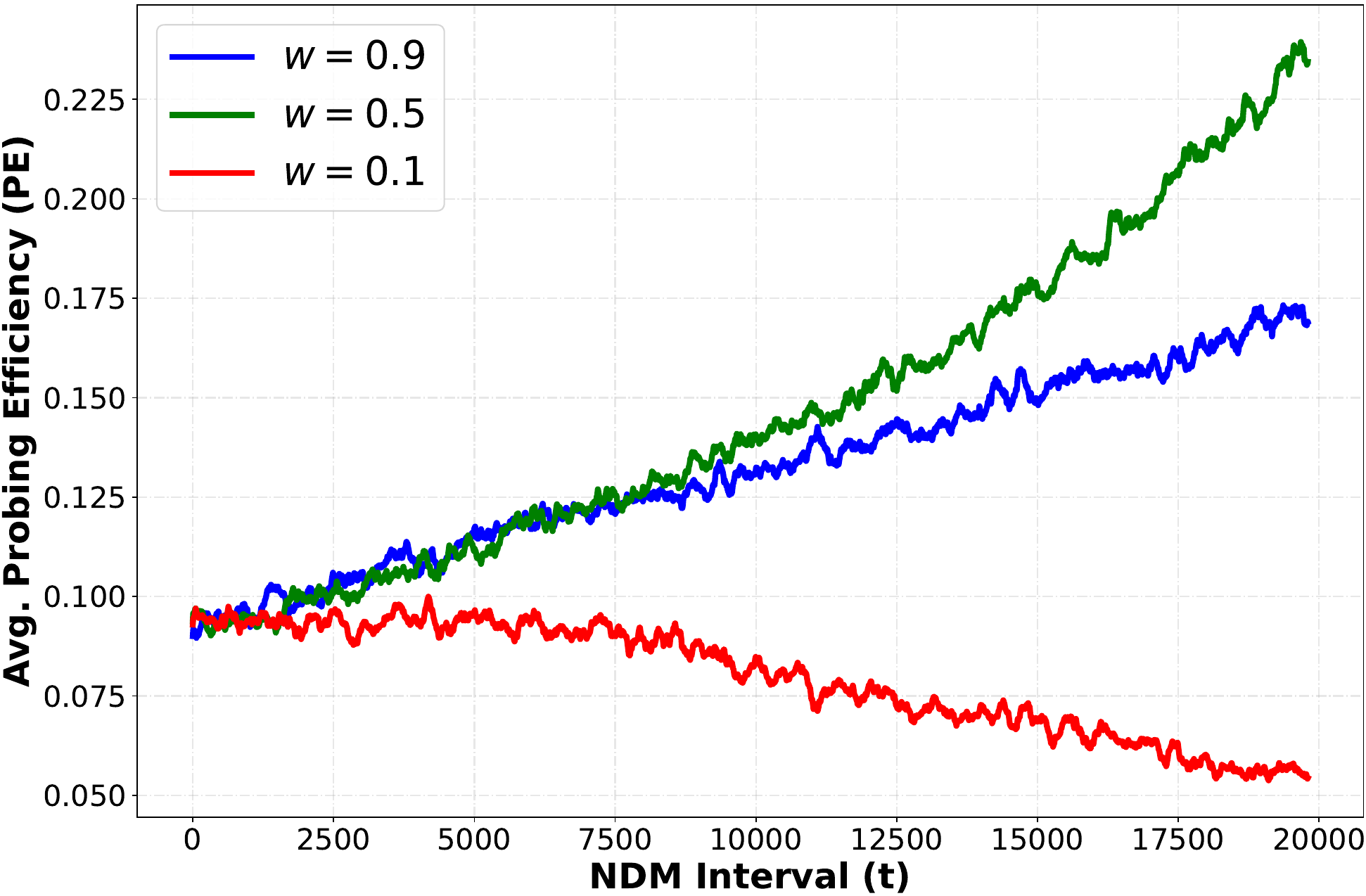}
        \captionsetup{format=hang}
        \caption{Probing efficiency (PE) over time}
        \label{dqn_pe_comp}
    \end{subfigure}
    \begin{subfigure}[t]{0.49\linewidth}
        \centering
        \includegraphics[width=0.975\linewidth]{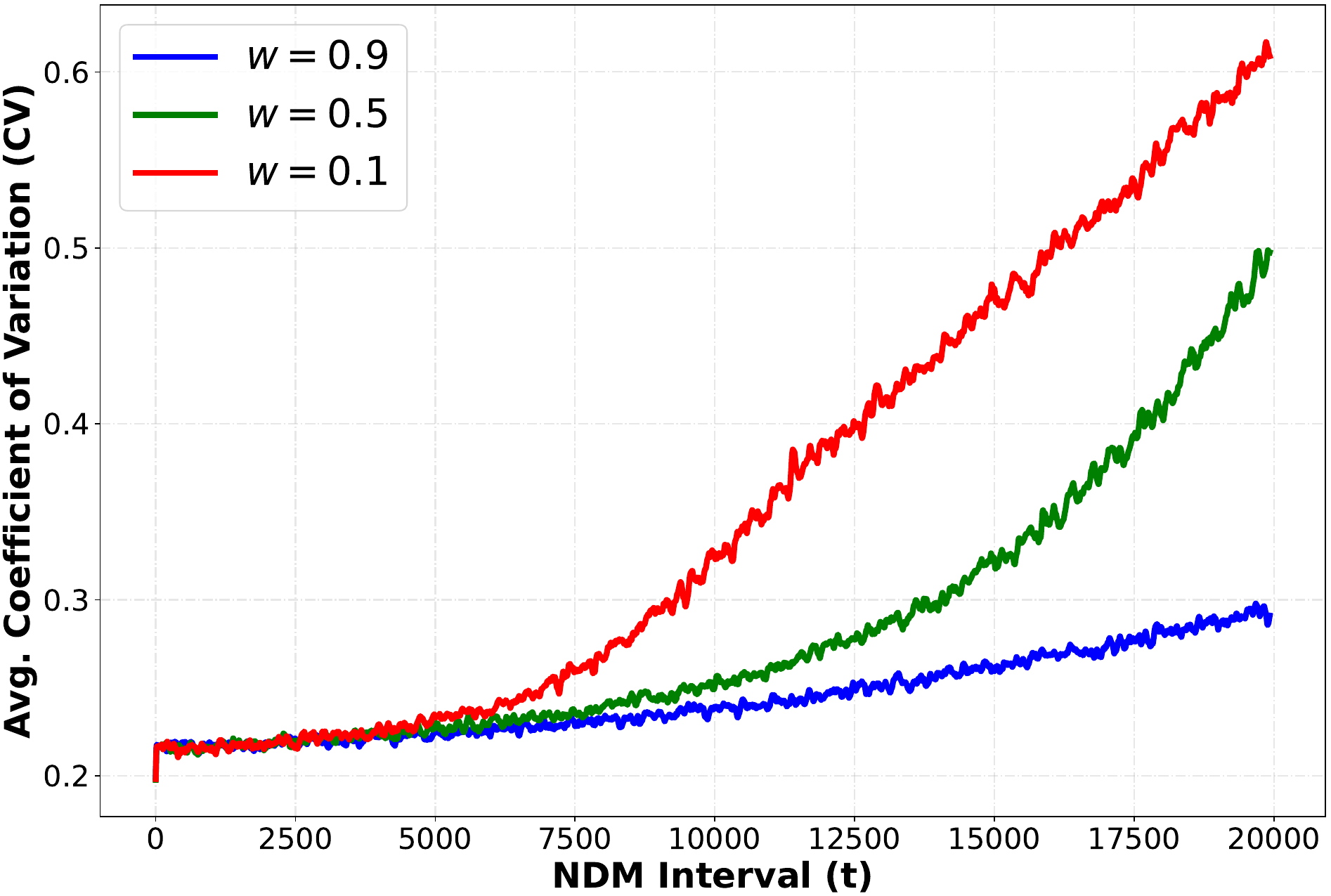}
        \captionsetup{format=hang}
        \caption{Coefficient of variation (CV) over time}
        \label{dqn_cov_comp}
    \end{subfigure}
    
    \begin{subfigure}[t]{0.49\linewidth}
        \centering
        \includegraphics[width=1\linewidth]{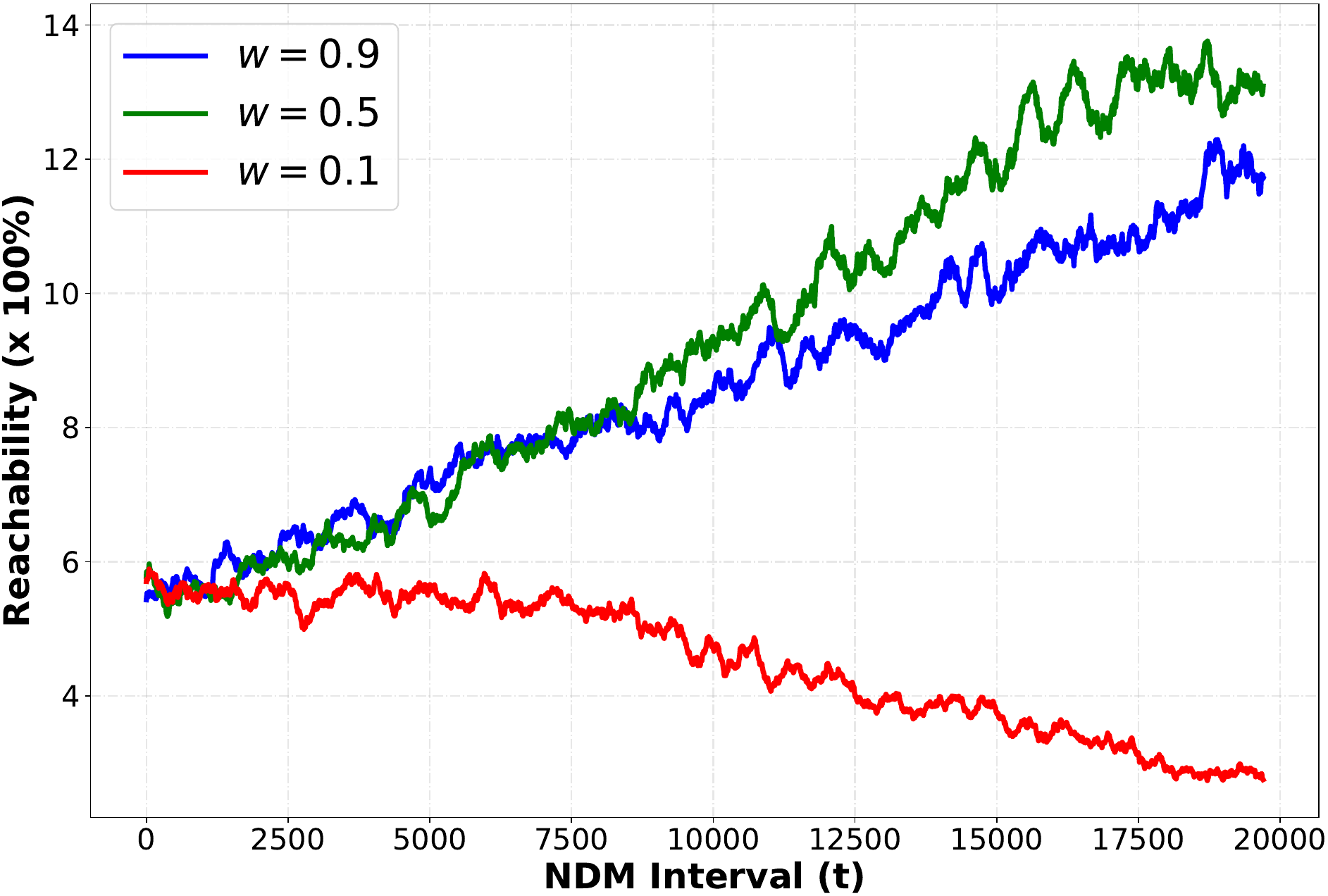}
        \captionsetup{format=hang}
        \caption{Reachability across the swarm over time}
        \label{dqn_reach_comp}
    \end{subfigure}
    \begin{subfigure}[t]{0.495\linewidth}
        \centering
        \includegraphics[width=1.0\linewidth]{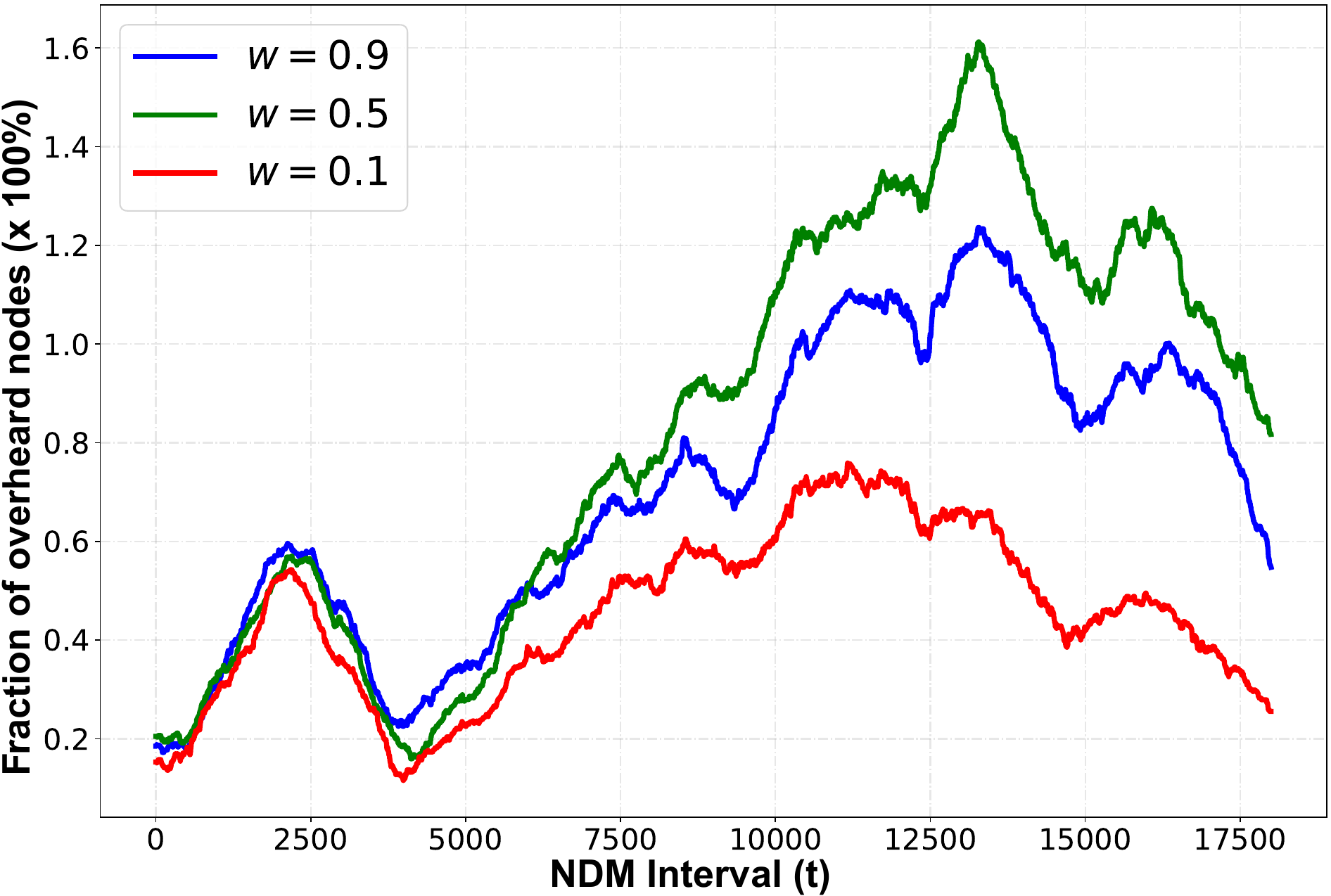}
        \captionsetup{format=hang}
        \caption{Fraction of nodes overheard by undesired users over time}
        \label{dqn_eve_comp}
    \end{subfigure}
    \captionsetup{format=hang}
    \caption{Performance comparison of DQN-based framework for different weights}
    \label{dqn_comp}
    \vspace{-3mm}
\end{figure}

\begin{figure}[t]
    \centering
    \begin{subfigure}[t]{0.49\linewidth}
        \centering
        \includegraphics[width=1\linewidth]{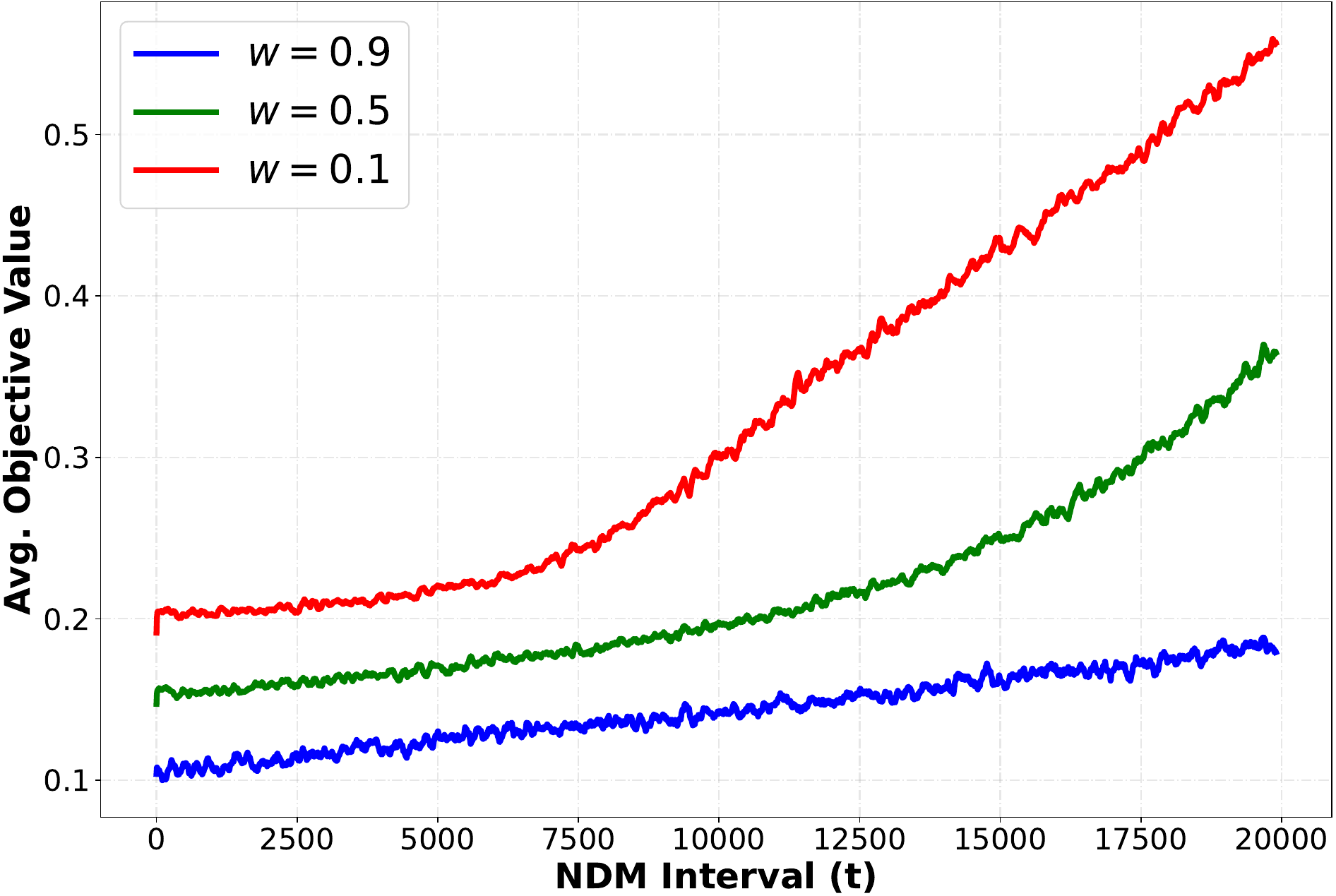}
        \captionsetup{format=hang}
        \caption{Average objective over time}
        \label{dqn_obj_comp}
    \end{subfigure}
    \begin{subfigure}[t]{0.49\linewidth}
        \centering
        \includegraphics[width=1\linewidth]{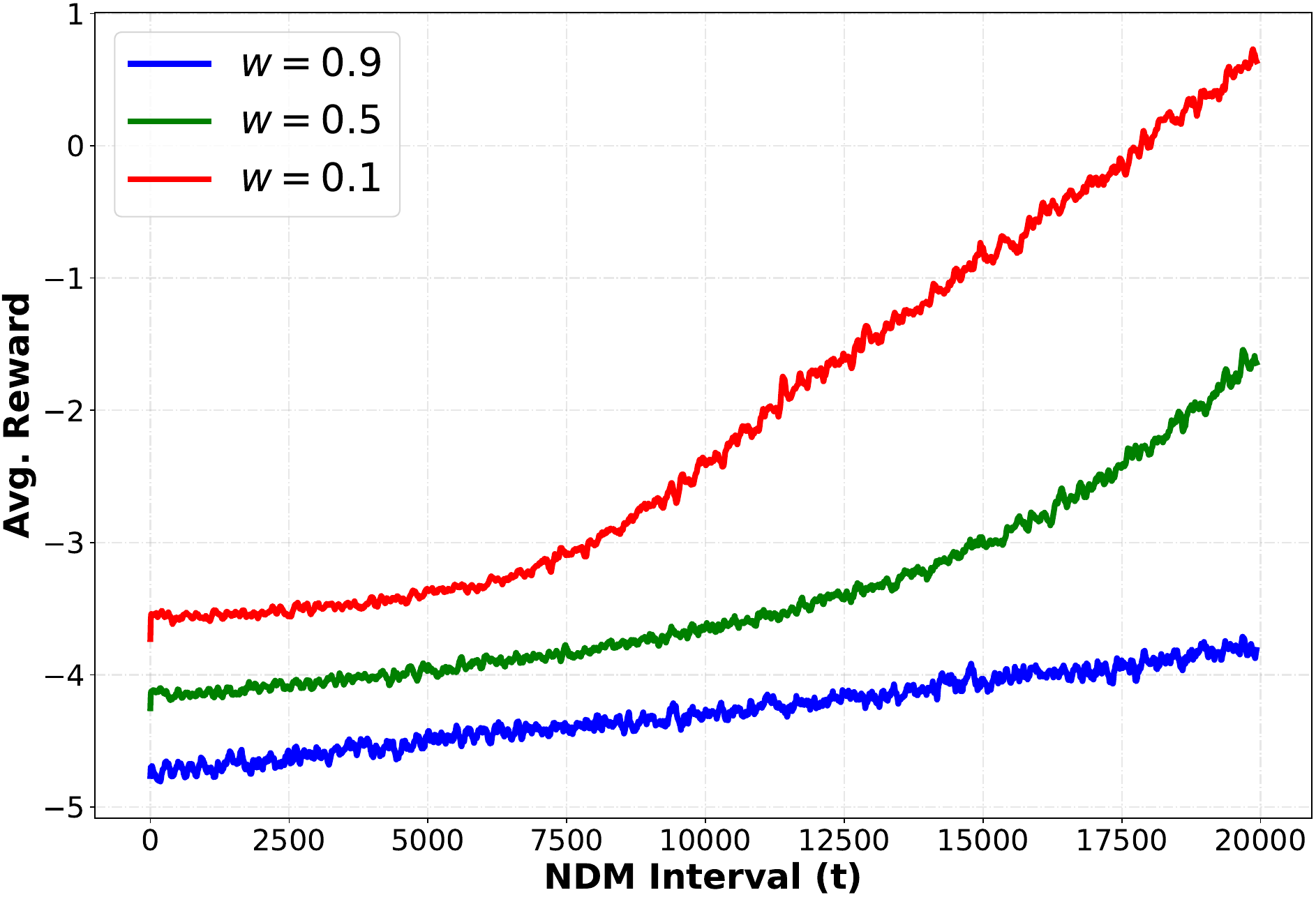}
        \captionsetup{format=hang}
        \caption{Average reward over time}
        \label{dqn_reward_comp}
    \end{subfigure}
    \captionsetup{format=hang}
    \caption{Average objective and reward trend of DQN agents for different weights}
    \label{dqn_obj_reward_comp}
    \vspace{-4mm}
\end{figure}

\section{Simulation Results and Discussion}

% \subsection{Simulation Environment}

% In the simulations, all the nodes are deployed in a $100\text{m}\times 100\text{m}$ grid. 

We develop a Python-based simulation environment where the nodes, each representing a UAV, are deployed in a $100\text{m}\times 100\text{m}$ grid. 
Both the transmission range ($R$) of directional nodes and the detection range ($R_d$) of undesired users are bounded by $30\text{m}$. 
% Each directional node is equipped with 8 on-board transceivers.
Other parameters are charted in Table \ref{tab:simulation-parameter}. 
% After repeating each simulation 50 times with varying seed values, the average statistics are studied for analysis.
%\subsection{Comparison with Baseline Algorithms}
We compare our DQN-driven framework with two baseline methods: Random and Q-Learning. In the Random method, a directional node randomly chooses the next probing direction. In the Q-Learning framework, we replace the DQN agents in our framework with Q-Learning agents using the same hyper-parameters as indicated in Table \ref{tab:dqn-parameter}. To keep the state-action space manageable for Q-Learning, we exclude the probing distribution $P_t$ from the state space while keeping the probing and outcome histories, $S_A$ and $S_o$.
%takes into account recent probing records to choose the next transceiver to probe.

\textit{DQN-based adaptive framework consistently outperforms the Random and Q-Learning baselines}, highlighting its ability to tune the probing patterns to meet the objective under constant topological shift.
Fig. \ref{alg_reach_comp} shows that DQN agents, particularly with higher weights (e.g., $w = 0.5, 0.9$) prioritizing fast discovery, quickly achieve and maintain high reachability.
% and ensures stable performance as time progresses. 
Simultaneously, DQN agents exhibit stability in ensuring secure communication, as shown in Fig. \ref{alg_eve_comp}. 
In comparison, Q-Learning shows slower adaptation, reflecting its difficulty to balance the trade-off under mobility. 
The Random baseline attains stability in reachability, but suffers the most from being overheard due to its non-adaptive probing behavior.

%\subsection{Influence of Different Weights, $w$} 

Fig. \ref{dqn_comp} reveals that \textit{higher $w$ accelerates PE and reachability, but causes broader exposure, resulting a higher risk of getting overheard by undesired users.} 
Higher $w$ allows directional nodes to explore the search space more by frequently switching sectors, which results in high PE and stable reachability, as illustrated in Figs. \ref{dqn_comp}a and \ref{dqn_comp}c. 
However, oftentimes excessive exploration pushes directional nodes to probe in directions outside the swarm layout, hurting the resulting PE, which is evident in the performance of $w=0.9$ in comparison to $w=0.5$. 
Conversely, lower $w$ prioritizes reusing a subset of on-board transceivers that, at the expense of low reachability, ensures reduced  exposure, higher CV, and consequently lower risk of getting overheard, as observed in Figs. \ref{dqn_comp}b and \ref{dqn_comp}d.

% The trend across Figs. \ref{dqn_comp}a-\ref{dqn_comp}d demonstrates that different weights allow DQN agents to autonomously tune their probing patterns to meet the requirement while operating under mobility.

Figs. \ref{dqn_comp}a and \ref{dqn_comp}b depict that with $w=0.1$, directional nodes attain high CV due to conservative scanning pattern, but the discovery rate gets severely punished. 
However, since the objective function in (\ref{eq:weighted_obj}) blends both PE and privacy, the CV gain at $w=0.1$ outweighs the drop in PE -- providing the highest net objective value as observed in Fig. \ref{dqn_obj_reward_comp}a. 
However, even after setting very high priority to discovery, directional nodes often face difficulty in finding neighbors in a consistent manner due to mobility. 
As a result, comparatively lower objective values are observed for $w=0.5$ and $w=0.9$.
Since the learning signal of a DQN agent is directly linked with objective values attained over time, this trend in Fig. \ref{dqn_obj_reward_comp}a eventually leads to securing relatively higher rewards for $w=0.1$ in comparison to other weights, as is evident from Fig. \ref{dqn_obj_reward_comp}b.
\vspace{-2mm}
\section{Conclusions}

A DQN-driven adaptive transceiver selection framework for mobile directional wireless systems is proposed. The framework can guide the transmitting node, acting as an independent DQN agent, toward striking a balance between network-wide connectivity and maintaining privacy in a mobile setting. A weighted mechanism is adopted to allow adaptive tuning of probing patterns based on network dynamics by using only recent local observations. A Python-based simulation environment is developed to study the influence of different weights $w$ on the performance of directional nodes deployed in a UAV swarm layout. The proposed DQN framework outperforms the Random and Q-Learning baselines. Weights that prioritize neighbor discovery lead to high PE and reachability, but increase the exposure in the search space. Conversely, lower weights can attain higher CV, which eventually acts as dominant factor to ensure higher net objective gain. This presents the weighting scheme of the framework as a tunable knob for balancing the connectivity-privacy trade-off. 

Future work can enhance the learning algorithm in a way that allows directional nodes to quantify the network reachability locally and ensure more active interaction with the environment to adapt their probing patterns. 
Additionally, there is a need to study the impact of providing nodes with partial global state information, such as total node count or network density, on the agents' learning behavior to maximize a global objective like network-wide connectivity.

% References
\bibliographystyle{IEEEtran}
\bibliography{refs}

\end{document}